\documentclass[conference]{IEEEtran}
\IEEEoverridecommandlockouts

\usepackage{cite}
\usepackage{amsmath,graphicx}
\usepackage{amssymb}
\usepackage{enumitem}
\usepackage{bm}
\usepackage{multicol,multirow,booktabs}
\usepackage{algorithm}
\usepackage{algpseudocode}
\usepackage{textcomp}
\usepackage{dsfont}
\usepackage[T1]{fontenc}
\usepackage[dvipsnames]{xcolor}
\usepackage{booktabs}

\usepackage[top=0.75in, left=0.63in, right=0.63in, bottom=1.08in]{geometry}
\def\BibTeX{{\rm B\kern-.05em{\sc i\kern-.025em b}\kern-.08em
    T\kern-.1667em\lower.7ex\hbox{E}\kern-.125emX}}
    
\begin{document}

\title{SNIPER Training: Single-Shot Sparse Training for Text-to-Speech}

\author{
    \thanks{This project has received funding from SUTD Kickstarter Initiative no. SKI 2021\_04\_06 and the President's Graduate Fellowship.}
    \IEEEauthorblockN{Perry Lam\IEEEauthorrefmark{1}, Huayun Zhang\IEEEauthorrefmark{2},
    Nancy F. Chen\IEEEauthorrefmark{2}, Berrak Sisman\IEEEauthorrefmark{3} and
    Dorien Herremans\IEEEauthorrefmark{1}
    }
    \IEEEauthorblockA{
        \IEEEauthorrefmark{1}Audio, Music, and AI Lab, Singapore University of Technology and Design, Singapore\\
        perry\_lam@mymail.sutd.edu.sg, dorien\_herremans@sutd.edu.sg
    }
    \IEEEauthorblockA{
        \IEEEauthorrefmark{2}Institute for Infocomm Research, A*STAR, Singapore\\
        \{Zhang\_Huayun, nfychen\}@i2r.a-star.edu.sg
    }
    \IEEEauthorblockA{
        \IEEEauthorrefmark{3}Speech \& Machine Learning Lab, The University of Texas at Dallas, USA\\
        Berrak.Sisman@UTDallas.edu
    }
}

\maketitle

%
\begin{abstract}
Text-to-speech (TTS) models have achieved remarkable naturalness in recent years, yet like most deep neural models, they have more parameters than necessary. Sparse TTS models can improve on dense models via pruning and extra retraining, or converge faster than dense models with some performance loss. Thus, we propose training TTS models using decaying sparsity, i.e. a high initial sparsity to accelerate training first, followed by a progressive rate reduction to obtain better eventual performance. This decremental approach differs from current methods of incrementing sparsity to a desired target, which costs significantly more time than dense training. We call our method SNIPER training: Single-shot Initialization Pruning Evolving-Rate training. Our experiments on FastSpeech2 show that we were able to obtain better losses in the first few training epochs with SNIPER, and that the final SNIPER-trained models outperformed constant-sparsity models and edged out dense models, with negligible difference in training time. Our code is available on Github\footnote{https://github.com/iamanigeeit/sniper}.
\end{abstract}


\section{Introduction}
\label{sec:intro}

 Although classical parametrized models only require $n$ parameters to fit $n$ data points,~\cite{bubeck} showed that large over-parameterized models with at least $nd$ parameters are in fact necessary for smooth data interpolation (where $d$ is the data dimensionality). This explains the dominance of ever-larger neural models, but we have now reached the era of extreme diminishing returns. For instance, \cite{limits} found that the compute power needed to reduce error rates by a factor of $x$ was at least $x^{10}$ across a range of image classification and natural language processing tasks. 

Counteracting this, researchers have proposed methods to compress models, such as knowledge distillation, tensor decomposition, quantization, and parameter sharing. The most flexible approaches involve sparsification techniques, which can be applied to various training stages and network architectures.



For instance, \cite{lottery-ticket} showed that when a dense network is initialized, some subnetworks (\textit{winning tickets}) can match or improve the same-iteration performance of the full network when trained in isolation. However, \cite{imp-fail} suggests that the winning tickets cannot be found without prior dense training and therefore requires much more time to train than the original unpruned model. The additional time taken holds for other sparsification schemes like dynamic pruning, which regrows weights during training according to gradients~\cite{nest} or momentum~\cite{sparse-momentum} in each backward pass. Sparse models also do not improve memory usage or training/inference time unless sparsities are very high (above 80\% for 2-D float32 tensors).

Since the goal is not just to produce a sparse model but to improve the performance-cost tradeoff, we propose a \textit{single-shot} sparse training scheme that progressively decreases sparsity to 0 rather than increasing it to a target sparsity. Intuitively, we want to direct gradient updates to more important weights at the start for faster convergence and to the less important weights later. We apply our method to FastSpeech2~\cite{fastspeech2} and evaluate its effectiveness via naturalness, intelligibility, prosody, and training time, comparing it to both dense and constant-sparsity models. We observe that the final SNIPER-trained models overtake the constant-sparsity models on nearly all metrics and surpass the dense model in most. 

In the next section, we introduce the related work on fast sparse training, followed by our proposed SNIPER training framework in Section \ref{sec:sniper}, our experiment setup in Section \ref{sec:experiments}, and finally we report our results in Section \ref{sec:results}.

\vspace{2pt}
\section{Related Work on Fast Sparse Training}
\label{sec:related}

\subsection{Hardware-Dependent Training}


There is a large body of research that proposes neural net accelerators on the hardware level~\cite{accelerators}, but such designs have not been commercially realized. The closest implementations currently available are NVIDIA Sparse Tensor Cores \cite{nvidia-tensor-cores}, where the inference speed doubles if 2:4 sparsity is satisfied in a tensor (i.e. every block of 4 elements has 2 zeros). \cite{zhou-and-ma} was the first to exploit this: they pruned the smallest weights from $\mathcal{W}$ to $\widetilde{\mathcal{W}}$ in each block before the forward pass. After the backward pass, they corrected the approximation error by changing the familiar weights update $\mathcal{W}_{t+1} = \mathcal{W}_t - \alpha_t g(\widetilde{\mathcal{W}}_t)$ to $\mathcal{W}_{t+1} = \mathcal{W}_t - \alpha_t (g(\widetilde{\mathcal{W}}_t) + \lambda (m_t \odot \mathcal{W}_t ))$ where $\alpha$ is the learning rate, $g$ is the gradient, $\lambda$ is a relative weight and $m$ is the weights mask. To allow sparse speedup on the backward pass as well,~\cite{hubara} proposed a greedy pruning method to find transposable masks that had the 2:4 property both row-wise and column-wise. They further proved that their method ensured that the $\ell_1$-norm of the pruned weights $W(P)$ was at most $< 2 W^*$ (weights from optimal pruning).

\subsection{Mixture of Experts}

Aside from hardware-specific methods, there has been a recent trend of creating models with multiple subnetworks and a routing layer to distribute inputs. This is a variant of \textit{Mixture-of-Experts} and is common in large language models. Pioneering this technique was the 1.6 trillion parameter Switch-Transformer model~\cite{switch-transformers}, which trained up to 4$\times$ faster than T5-XXL~\cite{t5}. While the original Transformer model contains a single dense feed-forward layer after each multi-head attention layer, the Switch-Transformer replaced it with 2,048 feed-forward blocks with a routing mechanism that decides which block should receive the attention output. In this way, the feed-forward blocks are sparse as most of them are not used, while training is accelerated as the data and feed-forward blocks can be parallelized. Ironically, this approach to sparse speedups makes the overall parameter count explode. 

Due to these shortcomings, there is still a largely unexplored space of hardware-independent sparse techniques that enhance training and maintain model size, which we attempt to address.

\section{SNIPER Training for TTS}
\label{sec:sniper}

\subsection{Sparsity in TTS}

While model compression~\cite{robust-distiller} and structured pruning~\cite{pruning-speech-rep} of speech representations have advanced in recent years, studies on sparsity in TTS are rare as TTS is generative in nature and harder to evaluate automatically, leading to ambiguity in proving technique effectiveness. Also, complex architectures impede component-wise characterization of sparse behaviour. 
Initial work on this topic~\cite{parp-tts} showed that both text-to-melspec and vocoder models could be pruned, while the output quality could be maintained or even improved. This was achieved by pruning the models (with zero weights allowed to be updated), training to convergence, and re-pruning. \cite{epic-tts} further tested five sparse techniques on a Tacotron2 baseline and demonstrated comparable performance when pruning before, during, or after training. Remarkably, single-shot network initialization pruning (SNIP) allowed the model to converge at least 1.9$\times$ faster at 40\% sparsity, although the quality degraded slightly.

\subsection{SNIP}

SNIP~\cite{snip} is one among many \textit{foresight pruning} methods that obtain a pruning mask before the training begins. It ranks the importance of a neural network weight by the change in loss if the weight is zeroed, estimated by taking the loss gradient with respect to the mask. For a binary mask $\bm{m}$ on weights $\bm{\mathrm{w}}$, the effect of removing $\bm{\mathrm{w}}_j$ on the loss $\mathcal{L}_j$ is
    $$\Delta \mathcal{L}_j(\bm{\mathrm{w}}; \mathcal{D}) \approx g_j(\bm{\mathrm{w}}; \mathcal{D})
    = \left. \frac{\partial \mathcal{L}(\bm{m} \odot \bm{\mathrm{w}}; \mathcal{D})}{\partial m_j} \right| _{\bm{m=1}}$$
    
Other prominent foresight pruning methods include (1) Gradient Signal Preservation~\cite{grasp}, a second-order extension of SNIP which calculates the gradient Hessian to compensate for correlated weights; (2) Neural Tangent Transfer~\cite{neural-tangent}, which finds a sparse network linearly approximating the training evolution of the dense network; and (3) SynFlow~\cite{synflow}, a data-free approach that iteratively prunes the weights with the lowest $\ell_1$-path norm. Further techniques involve comparing activation function output or weight magnitudes, possibly with lookahead to downstream layers~\cite{lamp}. Still,~\cite{noah-james} compared 13 of these pruning algorithms on image classification and found SNIP to be the most consistent performer over multiple sparsities and also relatively simpler to implement.

\subsection{SNIPER Training}

\begin{figure}[h!] \centering
  \includegraphics[width=0.9\columnwidth]{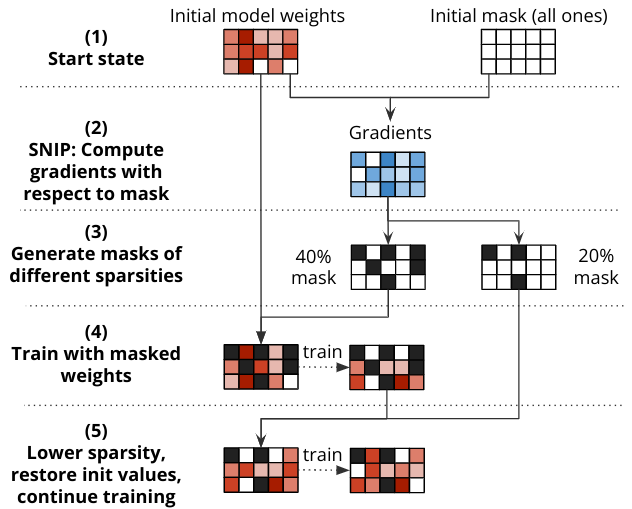}
  \caption{SNIPER example with 40\% sparsity reducing to 20\%.}
  \label{fig:sniper}
\end{figure}

Because of its consistency, we employ SNIP to determine the weights that should be masked at each sparsity level. We implement SNIP (Fig.~\ref{fig:sniper}) by saving the initial model weights, initializing the weight masks to $\mathbf{1}$ and computing the gradient with respect to the masks. The gradients are accumulated over the entire training dataset (or a subset of data) and saved. Once gradients are saved, we generate masks with different sparsities by simply removing the lowest absolute gradients; this takes very little time and the masks are reusable. The various masks can be swapped during training according to a given schedule, which constitutes the evolving-rate part of SNIPER. The formal process is given in Algorithm \ref{alg:sniper}.

We allow multiple options in addition to specifying the schedule: (1) custom scaling of learning rates by sparsity, (2) limiting the sparsity of parameters to prevent bottlenecks, (3) limiting the maximum parameter learning rate, (3) excluding parameters by name, (4) restoring newly activated weights during a sparsity reduction either to zeros or to the initial model values, (5) using a subset of the training corpus for gradient computation with a batch iterator, and (6) resuming from a previous state.

\newgeometry{top=0.75in, left=0.63in, right=0.63in, bottom=1.03in}

\newcommand{\D}{\mathcal{D}}
\newcommand{\Loss}{\mathcal{L}}
\newcommand{\PG}{\mathcal{P}}
\newcommand\g{\mathbf{g}}
\newcommand\m{\mathbf{m}}
\newcommand\w{\bm{\mathrm{w}}}
\newcommand\ind{\mathds{1}}

\begin{algorithm}[h!]
\caption{SNIPER Training}\label{alg:sniper}
\begin{algorithmic}
\Require Schedule $S$ [ epoch $t \rightarrow$ sparsity $s$ ], Dataset $\D$, Loss function $\Loss$, Epochs to train $t_{fin}$, Param groups $\PG$
\Ensure SNIP Gradients $\g$, Masks $M$ [ $t \rightarrow$ mask $\m$ ] \Ensure Final model weights $\w$
\State $\g \gets \mathbf{0}$
\State $\m \gets \mathbf{1}$
\State $\w \gets \text{XavierInit()}$ \Comment{Ensures consistent variance for SNIP}
\ForAll{batch $b \in \D$}
    \State $\m \gets \mathbf{1}$
    \State $\g \gets \g + \nabla_\m \Loss(\w; b)$
\EndFor
\State $n \gets \text{dim}(\mathbf{g})$
\ForAll{$t, s$ in $S$}  \Comment{$S[0]$ must be defined}
    \State $i \gets \lfloor ns \rfloor$
    \State $v \gets \text{sorted}(|\g|)[i]$
    \State $\m_j \gets \ind (|\g_j| > v) \ \ \forall j = 1 \dots \text{dim}(m)$
    \State $M[t] \gets \m$
\EndFor
\For{$t = 0, \dots, t_{fin}$}
    \If{$t$ in $M$}
        \State $\m \gets M[t]$
        \ForAll{$P \in \PG$} \Comment{Adjust LR of param groups}
            \State $\alpha_P \gets \text{dim}(P) / \sum_{j \in P} m_j$
        \EndFor
    \EndIf
    \ForAll{batch $b \in \D$}
        \State $\g' \gets \nabla_{\w} \Loss(\w \odot \m; b)$
        \ForAll{$P \in \PG$}
            \State $\w_P \gets \w_P - \alpha_P \cdot \g'_P$
        \EndFor
    \EndFor
\EndFor
\end{algorithmic}
\end{algorithm}

\section{Experiments}
\label{sec:experiments}

In our experiments, we compare SNIPER-trained models against both dense and constant-sparsity models in performance (naturalness and intelligibility) and training time.

\subsection{Baseline and Dataset}

We use the open-source ESPnet2 \cite{espnet2} version of FastSpeech2 with default settings and a pretrained ParallelWaveGAN~\cite{parallelwavegan} vocoder. 
The ground-truth durations were generated using the Montreal Forced Aligner (MFA) according to the original paper~\cite{fastspeech2}. We train FastSpeech2 for 400 epochs (320k steps) on the LJSpeech \cite{ljspeech} dataset with an 80:10:10 train-validation-eval split, retaining the model checkpoint with the best validation loss. In all experiments, the same initial values and random seed were used to reduce variability. 

\subsection{SNIPER}

In our SNIPER training runs, we exclude the embedding and normalization layers and set initial sparsity to 20\% and 40\% with a maximum individual parameter sparsity of 75\%. Thiscaused the overall model sparsity to be 19.8\% and 38.2\%). We then reduce sparsity to 0\% according to the schedules shown in Table~\ref{tab:comparisons}. These schedules were created by decreasing sparsity once the constant-sparsity training loss exceeds dense training loss at the same epoch. We compare them against the dense model (Baseline) and constant-sparsity models (Constant), as reported in Table \ref{tab:comparisons}. 

\addtolength{\skip\footins}{8pt}
\begin{table}[th]
    \centering
    \caption{Sparsity of models following \cite{epic-tts}. SNIPER schedules are similar to exponential step learning rate decay.}
    \label{tab:comparisons} \small
    \begin{tabular}{c|c|c}
    \toprule
    \textbf{Baseline}    & \textbf{Constant}     & \textbf{SNIPER}   \\ 
    \midrule
    \multirow{7}{*}{0\%} & \multirow{3}{*}{20\%} & 20\% (epoch 1-5)  \\
                         &                       & 10\% (epoch 6-10)                     \\
                         &                       & 0\% (epoch 11+)                       \\ 
    \cline{2-3}
                         & \multirow{4}{*}{40\%} & 40\%~(epoch 1-5)  \\
                         &                       & 20\% (epoch 6-10)                     \\
                         &                       & 10\% (epoch 11-20)                    \\
                         &                       & 0\% (epoch 21+)     \\
                         \bottomrule
    \end{tabular}
\end{table}



\section{Results}
\label{sec:results}

\subsection{Naturalness}

Naturalness was measured via mean opinion score (MOS) from 1 (worst) to 5 (best), conducted via PsyToolkit~\cite{psytoolkit-1,psytoolkit-2}. We obtained 19 full responses. As a control, participants also rated MOS for the human-generated ground truth samples (Natural). SNIPER training marginally gave the most consistent improvements over the baseline, as reported in Table \ref{tab:mos}.

\begin{table}[h]
    \caption{Resulting MOS scores. Bold = best model.} \small
    \label{tab:mos}
    \centering
    \begin{tabular}{c|c|c|c|c}
    \toprule
        \textbf{Sparsity}   & \textbf{Baseline}      &     {\textbf{Constant}} & {\textbf{SNIPER}}      &     \textbf{Natural}       \\
        \midrule
        20\%  & \multirow{2}{*}{3.76 \textpm 0.86} & 3.82 \textpm 0.94         & \textbf{3.83} \textpm 0.89         & \multirow{2}{*}{4.45 \textpm 0.64} \\
        40\%  &                                & 3.65 \textpm 0.95         & \textbf{3.78} \textpm 0.96         & \\
        \bottomrule
    \end{tabular}
\end{table}

Given that the mean MOS scores are close and the variability is large, we also performed a three-way preference test to enable a better comparison. Participants were asked to rate the best and worst samples in each set of 3 (Baseline, Constant, and SNIPER) samples at both 20\% and 40\% sparsities with 16 utterances each. 
The SNIPER-trained models were preferred in most settings, as reported in Table \ref{tab:preference}.

\begin{table}[h!] \centering
    \caption{Results for the preference test (\% rated best/worst). Bold = best result. The $p$-values are from Friedman's test with Baseline, Constant, and SNIPER as the 3 treatment groups ranked in order of (3-best, 2-middle, 1-worst).}
    \label{tab:preference}
   \small
    \begin{tabular}{l|c|c|c|c} \toprule
        \textbf{Sparsity}                 & \textbf{Baseline}         & \textbf{Constant}         & \textbf{SNIPER}                             & \textbf{$p$-value}        \\ 
        \midrule
        20\% (best)                       & \textbf{39.5}             & 23.0                      & 37.5                                        & \multirow{2}{*}{0.056}  \\
        20\% (worst) & 36.2 & 38.2 & \textbf{25.7} &                         \\ 
        \cline{5-5}
        40\% (best)                       & 36.2                      & 21.1                      & \textbf{42.8}                               & \multirow{2}{*}{0.101}  \\
        40\% (worst) & 34.2 & 34.2 & \textbf{31.6} &      \\
        \bottomrule
        \end{tabular}
\end{table}


\subsection{Intelligibility}

We used the commercial Speechmatics engine to transcribe generated audio across the entire test set for each model, and measured intelligibility by the word error rate (WER) of the transcripts against normalized LJSpeech text. On average, SNIPER-trained models were slightly better than baseline. 

\vspace{-4pt}

\begin{figure}[ht]
  \centering \small
  \includegraphics[width=0.9\linewidth]{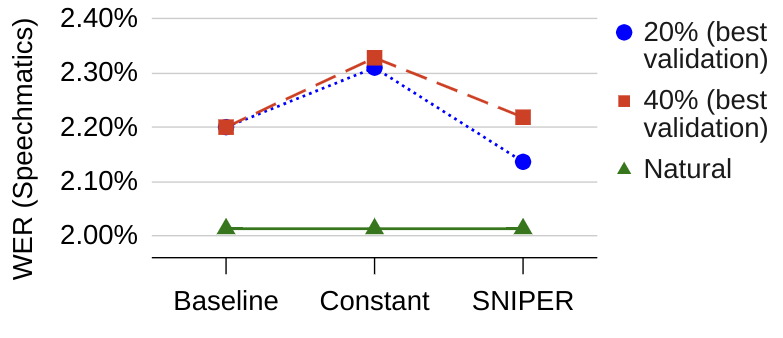}
  \setlength{\belowcaptionskip}{-10pt}
  \setlength{\abovecaptionskip}{-5pt}
  \caption{WER weighted by ground truth length.}
  \label{fig:wer}
\end{figure}



\vspace{-10pt}

\subsection{Training time}

All experiments were run on a single GeForce RTX3090 GPU. Calculating SNIP gradients across the whole dataset took 7.7 minutes and mask creation took less than 1 second (one-time operation). The slight delay in training sparse models comes from masking during the forward pass, but the difference for SNIPER is negligible.

\vspace{-6pt}
\begin{table}[h]
    \caption{Training time in hours.}
    \label{tab:complexity}
    \centering \small
    \begin{tabular}{c|c|c|c}
    \toprule
        {\textbf{Sparsity}} & \textbf{Baseline}        & \textbf{Constant}   & \textbf{SNIPER} \\
        \midrule
        20\%                & \multirow{2}{*}{13.1}    &
        13.5                & 13.1 \\
        40\%                &                          &
        13.5                & 13.2    \\
        \bottomrule
        \end{tabular}
\end{table}

\vspace{-4pt}
\section{Conclusion}

Our proposed SNIPER training method is the first attempt at using decreasing sparsity to improve TTS training. Even though the sparse FastSpeech2 models converged faster only during the initial training epochs, we have shown that the SNIPER-trained models are clearly superior to traditional sparse models and slightly better than dense models, especially in head-to-head comparisons. 
In future work, SNIPER may be expanded to allow automatic sparsity selection during training based on convergence rate, or improve upon dropout, which is a random and less directed form of sparsity.

\bibliographystyle{IEEEtran}
\bibliography{refs}
\end{document}